\documentclass[twocolumn]{amsproc}
\usepackage[margin=2cm]{geometry}
\usepackage{color}
\usepackage{tikz}
\usepackage{pgfplots}
\usepackage{undertilde}
\usepackage{fontenc}
\usepackage{yfonts}

\emergencystretch=\maxdimen
\begin{document}

\title{Reductions of lattice mKdV to $q$-$\mathrm{P}_{VI}$}
\author{Christopher M. Ormerod}
\address{Department of Mathematics and Statistics, Melbourne Campus,
La Trobe University,
Victoria 3086,
Australia}
\email{C.Ormerod@latrobe.edu.au}
%\thanks{Me}

\begin{abstract}
This Letter presents a reduction of the lattice modified Korteweg-de-Vries equation that gives rise to a $q$-analogue of the sixth Painlev\'e equation. This new approach allows us to give the first ultradiscrete Lax representation of an ultradiscrete analogue of the sixth Painlev\'e equation.
\end{abstract}

\maketitle

\section{Introduction}

This Letter will present a specific reduction of the non-autonomous lattice modified Korteweg-de-Vries equation \cite{nonalmkdv}, given by
\begin{equation}\label{lmkdv}
\alpha_l(w\bar{w} - \tilde{w}\tilde{\bar{w}})- \beta_m(w\tilde{w} - \tilde{w}\tilde{\bar{w}}) = 0
\end{equation}
where $w = w_{l,m}$, $\bar{w} = w_{l+1,m}$, $\tilde{w} = w_{l,m+1}$ and $\tilde{\bar{w}} = w_{l+1,m+1}$. The autonomous version of this equation is equivalent to $H3_{\delta = 0}$ in the list of multidimensionally consistent equations on quad-graphs \cite{AdlBobSur2003}. Reductions of \eqref{lmkdv} to $q$-analogues of the Painlev\'e equations were considered by Hay et al. \cite{Hay2007}. We wish to extend upon this work to provide a new way to think about reductions \cite{QuispelOrmerod}, demonstrating, as an example, how to obtain a $q$-analogue of the sixth Painlev\'e equation ($q$-$\mathrm{P}_{\mathrm{VI}}$) of Jimbo et al. \cite{Sakai:qP6}, given by
\begin{subequations}\label{qP6}
\begin{align}
f \hat{f}=\frac{q^2 \left(q^2 b_1 t^2+g a_2\right) \left(b_2 t^2+g a_1\right)}{\left(g b_1 q^2+a_2\right) \left(a_1+g
   b_2\right)},\\
g \hat{g}=\frac{\left(t^2 b_1 q^4+\hat{f} a_1\right) \left(t^2 b_2 q^4+\hat{f} a_2\right)}{q^2 \left(a_1+\hat{f}
   b_1\right) \left(a_2+\hat{f} b_2\right)},
\end{align}
\end{subequations}
as a reduction of \eqref{lmkdv}. Here we note that $\hat{t} = q^2 t$ for some fixed $q \in \mathbb{C}$ and the $a_i$ and $b_j$ are fixed parameters. This equation originally and a Lax representation first appeared as a connection-preserving deformation \cite{Sakai:qP6} and more recently as an equation governing a deformation of the little $q$-Jacobi polynomials \cite{orthoqP6}. While $q$-$\mathrm{P}_{\mathrm{VI}}$ has appeared as a reduction of a $q$-analog of the multi-component Kadomtsev-Petviashvili hierarchy \cite{KPqP6}, this is the first time that we know of that $q$-$\mathrm{P}_{\mathrm{VI}}$ has appeared as a reduction of a two-dimensional lattice equation. We will also obtain a new Lax pair by appealing to a new method developed in collaboration with Quispel \cite{QuispelOrmerod}.

We then show that this Lax representation may be ultradiscretized, hence, gives rise to a tropical Lax representation of an ultradiscrete analogue of the sixth Painlev\'e equation ($u$-$\mathrm{P}_{\mathrm{VI}}$) \cite{ultimate}, given by
\begin{subequations}\label{uP6}
\begin{align}
\hat{F} + F &= 2Q + \max(2Q + B_1 + 2T, G +A_2)\\ 
& + \max(B_2 + 2T, G+A_1) \nonumber\\ 
&- \max(G+B_1+2Q, A_2) \nonumber\\ &- \max(G + B_2, A_2), \nonumber\\
\hat{G} + G&= \max(2T+B_1+4Q,\hat{F}+A_1) \\
& + \max(2T+ B_2+4Q, \hat{f}+A_2) \nonumber\\ 
&-2Q - \max(A_1,\hat{F}+B_1) \nonumber \\ 
& - \max(A_2, \hat{F} + B_2) \nonumber,
\end{align}
\end{subequations}
where the $A_i$ and $B_j$ are fixed parameters in $\mathbb{R}$ and $\hat{T} = 2Q + T$ for some fixed $Q \in \mathbb{R}$. This is the first time that a tropical Lax representation of $u$-$\mathrm{P}_{\mathrm{VI}}$ has appeared that we know of.

This Letter is organized as follows. In Section \ref{sec:red} we will show that \eqref{qP6} arises as a reduction of \eqref{lmkdv}. In Section \ref{sec:Lax} we outline a new method of obtaining a Lax representation of a reduction to find a new Lax representation of \eqref{qP6}. In Section \ref{sec:qP3} we show how both the equation and Lax representation degenerate to give a $q$-analogue of the third Painlev\'e equation ($q$-$\mathrm{P}_{\mathrm{III}}$) and its Lax representation. Lastly, in Section \ref{sec:ud}, we show how the Lax representation may be ultradiscretized to give a tropical Lax representation of \eqref{uP6}.

\section{Reduction}\label{sec:red}
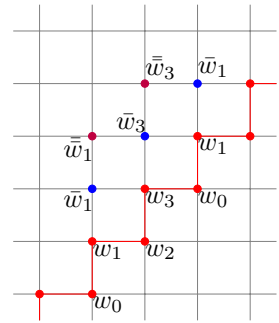
\begin{figure}[!ht]
\begin{tikzpicture}[scale=.7] 
 \draw[very thin,color=gray] (-1.5,-.5) grid (3.5,5.5);
    \draw[red] (-1,-.5) -- (-1,0) -- (0,0) -- (0,1) -- (1,1) -- (1,2) -- (2,2)-- (2,3) -- (3,3) -- (3,4) -- (3.5,4);
    \filldraw[red] (-1,0) circle(2pt);
    \filldraw[red] (0,0) circle(2pt);
    \filldraw[red] (0,1) circle(2pt);
    \filldraw[red] (1,1) circle(2pt);
    \filldraw[red] (1,2) circle(2pt);
    \filldraw[red] (2,2) circle(2pt);
    \filldraw[red] (2,3) circle(2pt);
    \filldraw[red] (3,3) circle(2pt);
    \filldraw[red] (3,4) circle(2pt);
    \begin{scope}[xshift = .3cm,yshift=-.2cm]
    \node at (0,0) {$w_0$};
    \node at (0,1) {$w_1$};
    \node at (1,1) {$w_2$};
    \node at (1,2) {$w_3$};
    \node at (2,2) {$w_0$};
    \node at (2,3) {$w_1$};    
    \end{scope}
    \node at (-.25,1.75) {$\bar{w}_1$};
    \node at (2.3,4.3) {$\bar{w}_1$};
    \node at (1.3,4.3) {$\bar{\bar{w}}_3$};
    \node at (-.25,2.75) {$\bar{\bar{w}}_1$};
    \node at (.75,3.25) {$\bar{w}_3$};
    \filldraw[blue] (0,2) circle(2pt);
    \filldraw[blue] (1,3) circle(2pt);
    \filldraw[blue] (2,4) circle(2pt);
    \filldraw[purple] (0,3) circle(2pt);
    \filldraw[purple] (1,4) circle(2pt);
\end{tikzpicture}
\label{figred}
\caption{The reduction and the labelling of variables.}
\end{figure}

We now consider the $(2,2)$-reduction, where we define $w_0$, $w_1$, $w_2$ and $w_3$ to be in a staircase, with $w_1$ directly above $w_0$, as in figure \ref{figred}. 

Notice that the evolution is consistent, so long as $\alpha_l/\beta_m = \alpha_{l+2}/\beta_{m+2}$, by which a separation of variables gives us
\begin{equation}\label{period}
\dfrac{\alpha_{l+2}}{\alpha_l} = \dfrac{\beta_{m+2}}{\beta_m} := q^2.
\end{equation}
To satisfy \eqref{period}, we define constants $a_i$ and $b_i$, for $i = 1,2$, by letting
\begin{eqnarray*}
\alpha_l = \left\{ \begin{array}{c p{2cm}} a_1 q^{l} & if $l$ is odd,\\
a_2 q^l & if $l$ is even,
\end{array}\right.\\
\beta_m = \left\{ \begin{array}{c p{2cm}} b_1 q^{m} & if $m$ is odd,\\
b_2 q^m & if $l$ is even.
\end{array}\right.
\end{eqnarray*}
If we let $t=q^{m-l}$, we have that $\beta_m/\alpha_l \propto t$, where the shift $m \to m+2$ is equivalent to $t \to q^2 t$. We solve \eqref{lmkdv} to find $\tilde{\tilde{w}}_0$ and $\tilde{\tilde{w}}_2$, given by
\begin{subequations}\label{weqs}
\begin{align}
\tilde{\tilde{w}}_0 &= \dfrac{w_2 \left(a_1 w_1+t b_2 w_3\right)}{t b_2 w_1+a_1 w_3},\\
\tilde{\tilde{w}}_2 &= \dfrac{w_0 \left(t b_1 w_1 q^2+a_2 w_3\right)}{t b_1 w_3 q^2+a_2 w_1},
\end{align}
where we may subsequently use the periodicity and \eqref{lmkdv} to obtain $\tilde{\tilde{w}}_0$ and $\tilde{\tilde{w}}_2$, given by
\begin{align}
\tilde{\tilde{w}}_1 &= \dfrac{w_3 \left(t b_1 \tilde{\tilde{w}}_2 q^2+a_1 \tilde{\tilde{w}}_0\right)}{t b_1 \tilde{\tilde{w}}_0
   q^2+a_1 \tilde{\tilde{w}}_2},\\
\tilde{\tilde{w}}_3 &= \dfrac{w_1 \left(t b_2 \tilde{\tilde{w}}_0 q^2+a_2 \tilde{\tilde{w}}_2\right)}{t
   b_2 \tilde{\tilde{w}}_2 q^2+a_2 \tilde{\tilde{w}}_0}.
\end{align}
\end{subequations}
Letting $w_0/w_2 = f/t$ and $w_1/w_3 = g/t$ gives \eqref{qP6}, where we now interpret $\tilde{\tilde{f}}$ and $\tilde{\tilde{g}}$ to be $\hat{f}$ and $\hat{g}$ respectively.

\section{Lax representation}\label{sec:Lax}

We use a different approach to reductions to that of Hay et al. \cite{Hay2007}. The general method will be further explored in a separate publication \cite{QuispelOrmerod}. 

We first note that \eqref{lmkdv} is multilinear and multidimensionally consistent, giving rise to the following Lax representation
\begin{subequations}
\begin{align}
\psi_{l+1,m} &= L_{l,m}\psi_{l,m},\\
\psi_{l,m+1} &= M_{l,m}\psi_{l,m},
\end{align}
\end{subequations}
where
\begin{align*}
L_{l,m} &= 
\begin{pmatrix}
 \dfrac{\gamma}{\alpha_l}  & \bar{w} \\
 \dfrac{1}{w} & \dfrac{\gamma  \bar{w}}{\alpha_l w} 
\end{pmatrix},\\
M_{l,m} &= \begin{pmatrix}
\dfrac{\gamma}{\beta_m}  & \tilde{w} \\
\dfrac{1}{w} & \dfrac{\gamma  \tilde{w}}{\beta_m w}
\end{pmatrix},
\end{align*}
and where $\gamma$ is a spectral parameter \cite{Hay2007}. 

We define two variables, 
\begin{align*}
x = \dfrac{q^{l}}{\gamma}, && t = q^{m-l},
\end{align*}
and a new linear system, $\phi(x,t)$, satisfying
\begin{subequations}\label{Lax}
\begin{align}
\label{LaxA}\phi(q^2x,t) &= A(x,t) \phi(x,t),\\
\label{LaxB}\phi(x,q^2t) &= B(x,t) \phi(x,t),
\end{align}
\end{subequations}
where operators $A(x,t)$ and $B(x,t)$ may be interpreted as operating in the $(2,2)$- and $(0,2)$-directions respectively in our original system. The $(2,2)$-operator has the effect of fixing $t$ and letting $z\to z/q^2$ and the $(0,2)$-operator fixes $z$ and lets $t \to q^2 t$. We may explicitly construct $A(x,t)$ and $B(x,t)$ in terms of $L$ and $M$:
\begin{align*}
A(x,t) &= L_{l+1,m+2}M_{l+1,m+1}L_{l,m+1}M_{l,m},\\
       &= \begin{pmatrix}
 \dfrac{1}{xa_2}  & w_0 \\
 \dfrac{1}{w_3} & \dfrac{w_0}{xa_2 w_3} 
\end{pmatrix}
\begin{pmatrix}
 \dfrac{1}{t x b_2}  & w_3 \\
 \dfrac{1}{w_2} & \dfrac{w_3}{t xb_2 w_2} 
\end{pmatrix}\\
&\begin{pmatrix}
 \dfrac{1}{xa_1}  & w_2 \\
 \dfrac{1}{w_1} & \dfrac{w_2}{xa_2 w_1} 
\end{pmatrix}
\begin{pmatrix}
 \dfrac{1}{t x b_1}  & w_1 \\
 \dfrac{1}{w_0} & \dfrac{w_1}{t xb_1 w_0} 
\end{pmatrix},\\
B(x,t) &= M_{l,m+1}M_{l,m},\\
&= \begin{pmatrix}
 \dfrac{1}{t x b_2}  & \tilde{\tilde{w}}_0 \\
 \dfrac{1}{w_1} & \dfrac{\tilde{\tilde{w}}_0}{t x b_2 w_1} 
\end{pmatrix}
\begin{pmatrix}
 \dfrac{1}{t x b_1}  & w_0 \\
 \dfrac{1}{w_0} & \dfrac{w_1}{txb_1 w_0} 
\end{pmatrix},
\end{align*}
where we have directly substituted for $x$ and $t$. The consistency of \eqref{LaxA} and \eqref{LaxB}, which reads
\begin{equation}\label{Comp}
A(x,q^2t)B(x,t) = B(q^2x,t)A(x,t),
\end{equation}
results in \eqref{weqs}. We may recast this system by the same identification that related \eqref{weqs} to \eqref{qP6}, with an additional factor, $h = w_3$, which we interpret to be a gauge factor \cite{Sakai:qP6}. Under this identification, we may manipulate the matrices to obtain an equivalent $A(x,t)$ in terms of $f$, $g$ and $h$;
\begin{align*}
A(x,t) &= \begin{pmatrix}
 \dfrac{1}{x a_2} & 1 \\
 \dfrac{1}{h} & \dfrac{1}{h x a_2}
\end{pmatrix}
\begin{pmatrix} 
\dfrac{1}{t x b_2} & h \\
\dfrac{f}{t} & \dfrac{f h}{t^2 x b_2}
\end{pmatrix}\\
&\begin{pmatrix}
 \dfrac{1}{x a_1} & \dfrac{t}{f} \\
 \dfrac{t}{g h} & \dfrac{t^2}{f g h x a_1}
\end{pmatrix}
\begin{pmatrix} 
\dfrac{1}{t x b_1} & \dfrac{g h}{t} \\
1 & \dfrac{g h}{t^2 x b_1}
\end{pmatrix},\\
B(x,t) &= \begin{pmatrix}
 \dfrac{1}{t x b_2} & \dfrac{b_2 t^2+g a_1}{t a_1+g t b_2} \\
 \dfrac{t}{g h} & \dfrac{b_2 t^2+g a_1}{g^2 h t x b_2^2+g h t x a_1 b_2}
\end{pmatrix}\\& \begin{pmatrix}
 \dfrac{1}{t x b_1} & \dfrac{g h}{t} \\
 \dfrac{t}{f} & \dfrac{g h}{f t x b_1}.
\end{pmatrix},
\end{align*}
where we have used the definitions and \eqref{weqs}. Note that
\begin{align*}
\det A(x,t) =& \left(\frac{1}{x^2 a_1^2}-1\right) \left(\frac{1}{x^2 a_2^2}-1\right) \\ 
&\left(\frac{1}{t^2 x^2 b_1^2}-1\right) \left(\frac{1}{t^2 x^2 b_2^2}-1\right),
\end{align*}
and that the leading matrices in the expansion of $A(x,t)$ around $x= 0$ and $x=\infty$ are both proportional to the identity matrix, meaning that \eqref{LaxA} and \eqref{LaxB} constitute a connection preserving deformation \cite{Sakai:qP6}. The compatibility, given by \eqref{Comp}, results in \eqref{qP6}. However, we obtain a necessary equation satisfied by the gauge factor:
\begin{equation}\label{guage}
\tilde{\tilde{h}} = h \dfrac{q^2 g \left(a_2+\tilde{\tilde{f}} b_2\right)}{t^2 b_2 q^4+\tilde{\tilde{f}} a_2},
\end{equation}
which bears some similarity to the equation satisfied by the gauge factor of Jimbo et al. \cite{Sakai:qP6}.

\section{Degeneration to $q$-$\mathrm{P}_{\mathrm{III}}$}\label{sec:qP3}

When $b_1 = b_2$, the evolution factorizes into two copies of the one mapping, which is also known as $q$-$\mathrm{P}_{\mathrm{III}}$, whose Lax representation was found by Papageorgiou et al. \cite{isomondromic}. Here, instead of having $m\to m+2$, we compute $m \to m+1$, where we have the equation
\begin{subequations}
\begin{align}
\label{w1}\tilde{w}_0 &= w_1 ,& \tilde{w}_2 &= w_3, \\ 
\label{w2}\tilde{w}_1 &= \tilde{\tilde{w}}_0  & \tilde{w}_3 &= \tilde{\tilde{w}}_2,
\end{align}
\end{subequations}
where we reuse \eqref{w1}. Using $w_0 = \utilde{w}$ and $w_1 = \utilde{w}$, the equation defining the evolution of $g$ is
\begin{equation}\label{qP3}
\tilde{g}\utilde{g} =\dfrac{\left(q^2 b_1 t^2+g a_2\right) \left(b_1 t^2+g a_1\right)}{\left(g b_1 q^2+a_2\right) \left(a_1+g b_1\right)},
\end{equation}
which a version of $q$-$\mathrm{P}_{\mathrm{III}}$ which is a known direct degeneration of $q$-$\mathrm{P}_{\mathrm{VI}}$ \cite{Sakai:qP6}. Under the identification above, we note that $\tilde{f} = q^2g$ and $b_1 = b_2$, hence we may also write $A(x,t)$ in terms of $g$ and $\utilde{g}$, however, in this case
\[
B(x,t) = M_{l,m} = \begin{pmatrix}
 \dfrac{1}{t x b_1} & \dfrac{g h}{t} \\
 \dfrac{t}{f} & \dfrac{g h}{f t x b_1}.
\end{pmatrix},
\]
where we replace \eqref{LaxB} with
\[
\phi(x,q t) = B(x,t)\phi(x,t),
\]
in which the new compatibility,
\[
A(x,qt)B(x,t) = B(q^2x,t)A(x,t) 
\]
gives \eqref{qP3}.

\section{Ultradiscretization}\label{sec:ud}

In this final section, we note that we may extend the above to a tropical Lax representation of $u$-$\mathrm{P}_{\mathrm{VI}}$. The ultradiscretization process successfully linked integrable cellular automata to discrete integrable systems \cite{ultra}. Given a rational subtraction-free function, $f(x_1,\ldots, x_2)$, we compute the ultradiscretization by introducing ultradiscrete variables, $X_i$, via the relation $x_i = e^{X_i/\epsilon}$. The ultradiscretization of $f$, $F$, is defined by
\[
F(X_1,\ldots,X_n) := \lim_{\epsilon \to 0} \epsilon \log f\left(x_1, \ldots, x_n\right).
\]
Applying this process to \eqref{qP6} gives \eqref{uP6}. The resulting system, \eqref{uP6}, is defined over the max-plus semifield, $S = \{\mathbb{R} \cup \{ -\infty\}, \otimes, \oplus\}$, where $a \times b = a+ b$ and $a \oplus b = \max(a,b)$, known as tropical multiplication and tropical addition respectively. We may extend these operations to matrices over $S$; if $U = (u_{i,j})$ and $V = (v_{i,j})$ are two matrices over $S$, then 
\begin{subequations}
\begin{align}
U \otimes V &:= \left( \max_{k}\left(u_{i,k} + v_{k,j}\right) \right),\\
U \otimes V &:= \left( \max\left(u_{i,j}, v_{i,j} \right) \right).
\end{align}
\end{subequations}
There are tropical Lax representations for partial difference equations \cite{Quispel:UDLaxs} and ultradiscrete Painlev\'e equations \cite{Corm1}. 

We note that the ultradiscrete analogue of the non-autonomous lattice modified Korteweg-de-Vries equation is given by 
\begin{align*}
\bar{W} - \tilde{W}=&\max( W + A_l, \tilde{\bar{W}}+ B_m)\\ 
&-\max( W + B_l, \tilde{\bar{W}}+ A_m),
\end{align*}
which we evolve by solving for $\bar{W}$ for each quadrilateral given a well-posed Cauchy problem. This system admits the tropical Lax representation
\begin{subequations}
\begin{align}
\Psi_{l+1,m} &= \mathcal L_{l,m} \otimes \Psi_{l,m},\\
\Psi_{l,m+1} &= \mathcal M_{l,m} \otimes \Psi_{l,m},
\end{align}
\end{subequations}
where
\begin{align*}
\mathcal L_{l,m}(z) &= 
\begin{pmatrix}
 \Gamma - A_l  & \bar{W} \\
 -W & \Gamma + \bar{W}- A_l -W
\end{pmatrix}\\
\mathcal M_{l,m}(z) &= \begin{pmatrix}
\Gamma - B_m  & \tilde{W} \\
-W & \Gamma + \tilde{W}- B_m -W
\end{pmatrix}.
\end{align*}
The compatibility condition reads
\[
\mathcal L_{l,m+1} \otimes \mathcal M_{l,m} = \mathcal M_{l+1,m} \otimes \mathcal L_{l,m}.
\]
This is the non-autonomous counterpart of the Lax pair of Quispel et al. \cite{Quispel:UDLaxs}. The $(2,2)$-reduction in the ultradiscretized variables, $W_i$, is given by 
\begin{subequations}\label{udweqs}
\begin{align}
\tilde{\tilde{W}}_0 -W_2 &= \max(A_1 + W_1, T+ B_2 + W_3)\\ 
& - \max(T+ B_2 + W_1, A_1 + W_3),\nonumber \\
\tilde{\tilde{W}}_2-W_0 &= \max(T+ B_1 + W_1 + 2Q, A_2 + W_3)\\ 
& - \max(T + B_1+ W_3+2Q,A_2+W_1), \nonumber \\ 
\tilde{\tilde{W}}_1 - W_3 &= \max(T+B_1+ \tilde{\tilde{W}}_2 +2Q, A_1 + \tilde{\tilde{W}}_0)\\
 &- \max(T+B_1+2Q+\tilde{\tilde{W}}_0,A_1 + \tilde{\tilde{W}}_2),\nonumber \\
\tilde{\tilde{W}}_3- W_1 &= \max(T+ B_2 + \tilde{\tilde{W}}_0, A_2 + \tilde{\tilde{W}}_2)\\
& - \max(T+B_2 + \tilde{\tilde{W}}_2,A_2 + \tilde{\tilde{W}}_0).\nonumber 
\end{align}
\end{subequations}
It should be clear that $F = T+ W_0 - W_2$ and $G = T + W_1  -W_3$ satisfies \eqref{uP6}.

The Lax representation follows in a similar manner, as we define the tropical linear system 
\begin{subequations}\label{UDLax}
\begin{align}
\label{UDLaxA}\Phi(2Q + X,T) &= \mathcal A(X,T) \otimes \Phi(X,T),\\
\label{UDLaxB}\Phi(X,2Q+ T) &= \mathcal B(X,T) \otimes \Phi(X,T),
\end{align}
\end{subequations}
where 
\begin{align*}
\mathcal A(X,T) &= \begin{pmatrix}
-X-A_2 & 0 \\
-H & -H-X-A_2
\end{pmatrix}\\
& \otimes
\begin{pmatrix} 
-T-X-B_2 & H \\
F-T & F+H - 2T-X-B_2
\end{pmatrix}\\
&\otimes\begin{pmatrix}
-X-A_1 & T-F \\
 T-G-H & 2T - F-G-H-X-A_1
\end{pmatrix}\\
&\otimes \begin{pmatrix} 
-T-X-B_1 & G+H-T \\
0 & G+H - 2T-X-B_1
\end{pmatrix},\\
\mathcal B(X,T) &= \left( \begin{array}{c p{3cm}}
-T-X-B_2  & $F+\tilde{\tilde{W}}_0 - W_0-T$ \\
 T-G-H & $F+ \tilde{\tilde{W}}_0-W_0 - T-X-G-H-B_2$
\end{array}\right)\\& \otimes \left( \begin{array}{c p{2cm}}
-T-X-B_1  & $G+H-T$ \\
 T-F & $G+H-F-T-X-B_1$
\end{array}\right),
\end{align*}
where 
\begin{align*}
\tilde{\tilde{W}}_0 -W_0 &= \max(G+A_1,2T+B_2) \\
&- \max(F+A_1,F+G+B_2).
\end{align*}
This linear problem defines an ultradiscrete connection preserving deformation \cite{Corm2}. The resulting compatibility, 
\[
\mathcal A(X,2Q+T) \otimes \mathcal B(X,T) = \mathcal B(2Q+X,T) \otimes \mathcal A(X,T)
\]
gives \eqref{uP6} by identifying $\hat{F}$ and $\hat{H}$ with $\tilde{\tilde{F}}$ and $\tilde{\tilde{G}}$ respectively. There addition equation for the gauge factor reads
\begin{align*}
\tilde{\tilde{H}}- H =& 2Q + G + \max(A_2,\tilde{\tilde{F}} + B_2) \\
&- \max(2T+B_2+4Q,\tilde{\tilde{F}}+A_2).
\end{align*}
The same degeneration, when $B_1 = B_2$, gives both $u$-$\mathrm{P}_{\mathrm{III}}$, given by
\begin{align}
\label{uP3}\tilde{G} + \utilde{G} =& \max( 2Q + B_1+2T,G+A_2) \\
 &+ \max(B_1+2T, G + A_1)\nonumber\\
 & - \max(G+B_1 + 2Q, A_2)\nonumber\\
 & - \max(G+B_1+2Q,A_2)\nonumber,
\end{align}
and its Lax representation.

\section{Conclusion}

We have uncovered a relation between the lattice modified Korteweg-de-Vries equation and a $q$-analogue of the sixth Painlev\'e equation. A remarkable consequence of the derivation of the Lax representation is that the matrices defined by the associated linear problem factorize astonishingly into linear factors in the spectral variable. This factorization is apparent through the method in which these matrices were derived \cite{QuispelOrmerod}. 

\section{Acknowledgments} 

The author would like to thank G. R. W. Quispel and S. Lobb for their contributions. This research is supported by Australian Research Council Discovery Grant \#DP110100077.

\end{document}